\documentclass{article}
\usepackage{spconf,amsmath,graphicx,hyperref}
\usepackage{amssymb}
\usepackage{algorithm}
\usepackage{algpseudocode}
\usepackage{booktabs}
\title{DDSA: Dual-Domain Strategic Attack for Spatial-Temporal Efficiency in Adversarial Robustness Testing}
\name{Jinwei Hu\textsuperscript{1},
Shiyuan Meng\textsuperscript{2},
Yi Dong\textsuperscript{1*}\thanks{* Corresponding author: yi.dong@liverpool.ac.uk},
Xiaowei Huang\textsuperscript{1}}

\address{\textsuperscript{1} University of Liverpool, Liverpool, UK \\
         \textsuperscript{2} Shanghai Artificial Intelligence Laboratory, Shanghai, China}

\begin{document}
\ninept
\maketitle
\begin{abstract}
Image transmission and processing systems in resource-critical applications face significant challenges from adversarial perturbations that compromise mission-specific object classification. Current robustness testing methods require excessive computational resources through exhaustive frame-by-frame processing and full-image perturbations, proving impractical for large-scale deployments where massive image streams demand immediate processing. This paper presents DDSA (Dual-Domain Strategic Attack), a resource-efficient adversarial robustness testing framework that optimizes testing through temporal selectivity and spatial precision. We introduce a scenario-aware trigger function that identifies critical frames requiring robustness evaluation based on class priority and model uncertainty, and employ explainable AI techniques to locate influential pixel regions for targeted perturbation. Our dual-domain approach achieves substantial temporal-spatial resource conservation while maintaining attack effectiveness. The framework enables practical deployment of comprehensive adversarial robustness testing in resource-constrained real-time applications where computational efficiency directly impacts mission success.
\end{abstract}

\section{Introduction}
Real-time image transmission and processing systems have become indispensable in real-world large-scale applications, particularly in search-and-rescue (SAR) operations, agricultural monitoring systems, and social media image classification \cite{lygouras2019unsupervised, martinez2021search,Huang_2025_CVPR}. These applications can generate massive image streams requiring immediate processing: for instance, unmanned aerial vehicles (UAVs) capture hundreds of images per minute during missions \cite{8682048}, while social media platforms process millions images for real-time classification \cite{alam2018processing}. These systems rely heavily on robust computer vision algorithms, where classification failures of critical image can have negative impacts \cite{10330036}. However, recent research has revealed even state-of-the-art deep learning models, despite achieving remarkable accuracy in controlled environments, exhibit concerning vulnerability to carefully crafted adversarial examples (AEs) \cite{szegedy2013intriguing,11077439}. These perturbations, often imperceptible to human, can cause well-trained neural networks to produce catastrophically incorrect predictions \cite{carlini2017evaluating, 9525540}, making adversarial robustness testing essential for ensuring system reliability in real-world deployment.

\begin{figure}[htbp]
\centering
\vspace{-0.1cm}
\includegraphics[width=0.4\textwidth]{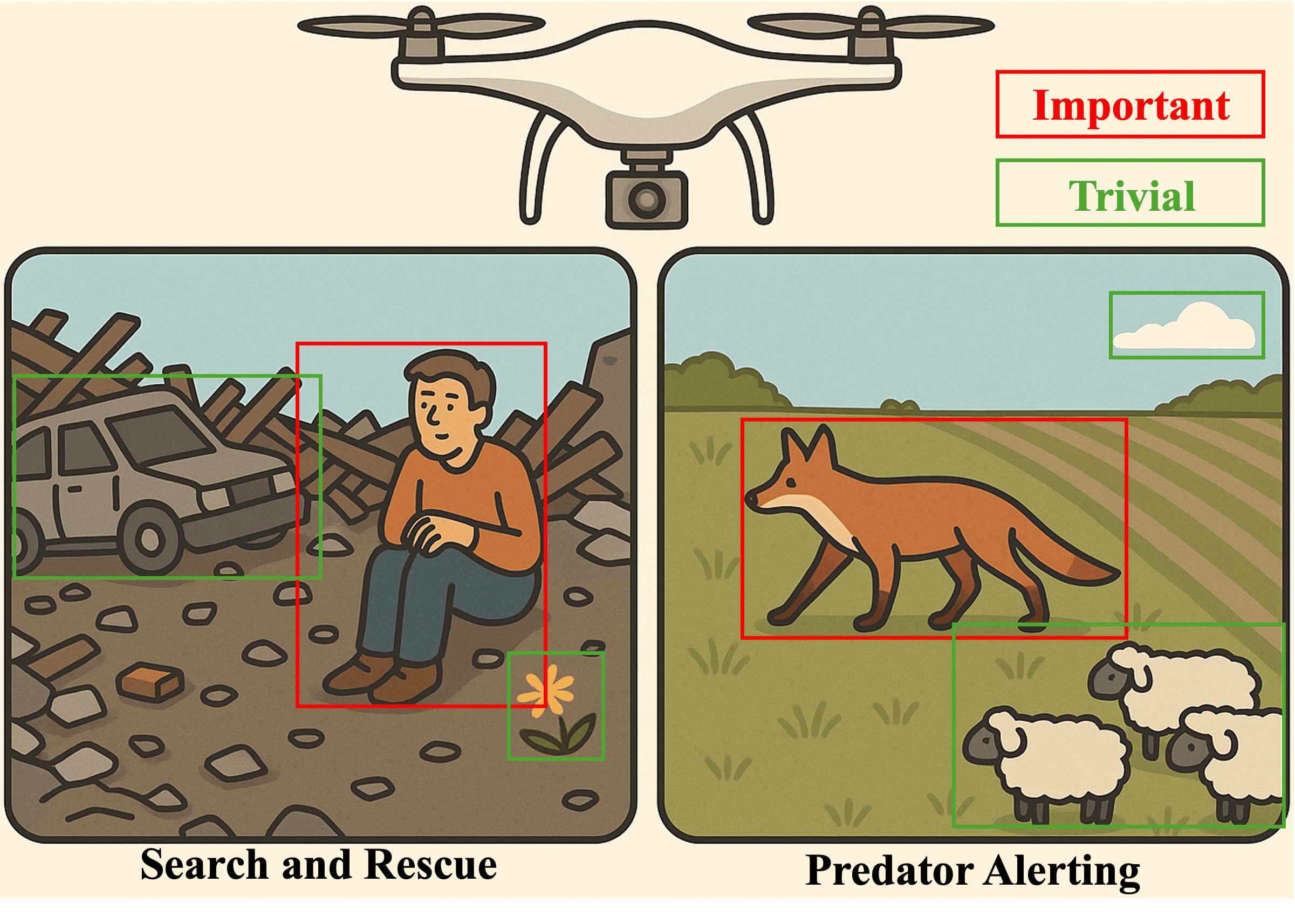}
\vspace{-0.3cm}
\caption{Task-dependent classification priority in image signal processing systems.}
\label{fig: motivation}
\end{figure}

Traditional adversarial robustness testing typically demands substantial computational resources, relying on exhaustive image-by-image processing and full-image adversarial attacks like Fast Gradient Sign Method (FGSM) and Projected Gradient Descent (PGD) \cite{https://doi.org/10.48550/arxiv.1412.6572, Zheng_Chen_Ren_2019, carlini2017evaluating}. While these methods are effective, they are impractical for large-scale resource-constrained scenarios where not all content requires equal testing priority \cite{machado2021adversarial}. As shown in Fig. \ref{fig: motivation}, In SAR operations, human classification is mission-critical where misclassification could have life-threatening consequences, while vegetation or vehicles identification has minimal impact on rescue success. In agricultural predator alerting systems, predator classification demands immediate response to protect livestock, whereas other features can tolerate misclassifications. This content heterogeneity motivates the need for resource-efficient adversarial robustness testing that prioritizes critical object classes.
% \vspace{-0.45cm}

To address these gaps, this paper presents a novel efficient Dual-Domain Strategic Attack (DDSA) adversarial testing framework through temporal selectivity and spatial precision. Our approach introduces two key components: a scenario-aware temporal trigger function that identifies critical content requiring robustness evaluation based on class priority and model uncertainty, thereby reducing meaningless overhead in sequential processing, and an explainable AI-guided spatial targeting mechanism that precisely locates the most influential pixel regions using Integrated Gradients for focused perturbation testing. This dual-domain strategy achieves significant resource conservation while maintaining attack effectiveness for priority classes, enabling practical deployment in resource-constrained safety-critical applications where computational efficiency and testing coverage are paramount.

\section{Related Work}
While significant advances have been achieved in developing sophisticated adversarial attack algorithms, efforts aimed at enhancing their resource efficiency for large-scale image processing applications lag noticeably behind \cite{Wang_2021_CVPR,kurakin1607adversarial,feinman2017detecting}. In large-scale image processing scenarios, not every image carries equal significance for robustness testing, and within individual images, not every region contributes meaningfully to critical decision-making processes. Regarding adversarial robustness testing efficiency, our work is intricately connected to two primary research directions. The first involves temporal processing challenges in sequential image streams, where traditional adversarial attack methodologies apply exhaustive image-by-image processing without discriminating between mission-critical and routine content, leading to substantial computational waste on non-essential images \cite{chen2021appending}. Secondly, from a single-image perspective, we observe that spatial regions contribute unevenly to model predictions. Accordingly, we adopt explainable AI (XAI) techniques that have been widely applied across diverse domains \cite{HU2024148465,XIE2024110564} for feature attribution \cite{simonyan2013deep,sundararajan2017axiomatic,shrikumar2017learning}, enabling us to identify and target spatially critical regions rather than performing indiscriminate full-image perturbations. Distinguishing our work from these established paths, our approach address both temporal content heterogeneity and spatial region significance through intelligently determining \textit{when to attack} (temporal selectivity) and \textit{where to attack} (spatial targeting) for resource-efficient adversarial robustness testing in large-scale deployment scenarios.

% \vspace{-0.15cm}
\section{Problem Formulation}
This paper addresses resource efficiency challenges in adversarial robustness testing across two key dimensions. In the temporal dimension, sequential image processing often encounters frames with minimal object variation or contains objects with varying safety criticality---for instance, in search-and-rescue operations, human detection errors can have life-threatening consequences, while misclassified debris or vegetation poses significantly lower risks. Applying uniform attack algorithms to every frame results in substantial computational waste on inconsequential content \cite{9638636}. In the spatial dimension, indiscriminate pixel-level perturbations across entire images prove equally inefficient, as specific regions such as edges or background areas contribute minimally to model predictions while potentially introducing unnecessary image distortion. To quantify this dual-domain efficiency problem, we define a safety-related evaluation metric that balances attack success against resource utilization. The total performance score is defined as:
\begin{equation}
    P_{\text{total}} = \sum_{i=1}^{N_X} P(x_i) + \sum_{j=1}^{N_D} P(d_j)
\end{equation}
where $X$ represents the list of images with labels belonging to the designated critical list, and $D$ represents the collection of images with labels outside this list. $N_X$ and $N_D$ are the total number of images in $X$ and $D$. The performance scores are defined as:
\begin{equation}
    P(y) =
\begin{cases}
\alpha, & \text{if } y \in X \text{ and attack is successful} \\
\beta, & \text{if } y \in X \text{ and attack is unsuccessful} \\
\gamma, & \text{if } y \in D \text{ and attack is attempted}
\end{cases}
\end{equation}
where $\alpha$, $\beta$, and $\gamma$ are reward coefficients with $\beta < 0$ and $\gamma < 0$, reflecting the penalties for unsuccessful attacks and resource waste. This formulation establishes the foundation for our dual-domain framework that determines both when to attack and where to attack.

\section{Dual-Domain Strategic Attack}

The DDSA framework integrates temporal selectivity and spatial precision for efficient adversarial robustness testing in image classification systems through three interconnected steps: temporal trigger function, spatial feature location, and strategic attack generation.
% \vspace{-0.35cm}
\subsection{Temporal Trigger Function}
The trigger function determines attack timing by evaluating the priority and uncertainty of the predicted class for each input image. For image $x_i$ with predicted class $\hat{y}_i$, the classification priority score is:
\begin{equation}
\label{q_function}
Q_i = \delta \cdot R^{scena}_{\hat{y}_i} + \zeta \cdot R^{data}_{\hat{y}_i} + \theta \cdot R^{consist}_{\hat{y}_i}
\end{equation}
where $\delta$, $\zeta$, and $\theta$ are relative importance factors. The scenario component $R^{scena}_{\hat{y}_i}$ incorporates domain expertise, assigning higher priorities to mission-critical classes (e.g., human classification in SAR operations receives maximum priority while background receives minimal priority). Moreover, the data-driven component quantifies the classification quality and prediction strength:
\begin{equation}
R^{data}_{\hat{y}_i} = \omega_1 \cdot \text{Conf}(\hat{y}_i) + \omega_2 \cdot \left(1 - \frac{\text{Entropy}(\mathbf{p}_i)}{\log C}\right)
\end{equation}
where $\text{Conf}(\hat{y}_i) = \max(\mathbf{p}_i)$ is the predicted class confidence, $\text{Entropy}(\mathbf{p}_i) = -\sum_{k=1}^{C} p_{i,k} \log p_{i,k}$ measures the prediction uncertainty across all $C$ classes, and $\mathbf{p}_i$ is the softmax output probability vector. The normalized entropy term $(1 - \text{Entropy}(\mathbf{p}_i)/\log C)$ ensures higher scores for more confident predictions. 

The consistency component employs Bayesian uncertainty estimation via Monte Carlo dropout to assess model reliability:
\begin{equation}
R^{consist}_{\hat{y}_i} = 1 - \sqrt{\frac{1}{T}\sum_{t=1}^{T} (P^t(\hat{y}_i|x_i))^2 - \left(\frac{1}{T}\sum_{t=1}^{T} P^t(\hat{y}_i|x_i)\right)^2}
\end{equation}
where $T$ denotes the number of dropout runs, $P^t(\hat{y}_i|x_i)$ represents the predicted probability for class $\hat{y}_i$ during the $t$-th run, and the square root term computes the standard deviation of predictions. A lower standard deviation results in a higher consistency score. 

Attack 
initiation then 
occurs when $Q_i > \tau$ for a predefined threshold $\tau$, ensuring that adversarial robustness testing focuses on high-priority, confident predictions where classification failures would have significant consequences.

\subsection{Spatial Feature Location}
Spatial targeting leverages Integrated Gradients (IG) \cite{sundararajan2017axiomatic,goh2021understanding} to identify critical pixel regions for efficient adversarial perturbation. For input $x_i$ and baseline $x'$ (blank image), the integrated gradient is:
c
where $\odot$ denotes element-wise multiplication. The integrated gradients $G(x_i)$ are reshaped to produce a saliency map $L_i \in \mathbb{R}^{H \times W}$ that highlights pixels most influential for classification decisions. The binary attack mask is generated via importance thresholding:
\begin{equation}
\label{s3}
B_i(h,w) = \mathbb{I}[L_i(h,w) \geq t] = \begin{cases} 1, & \text{if } L_i(h,w) \geq t \\ 0, & \text{otherwise} \end{cases}
\end{equation}
where $L_i(h,w)$ denotes the importance score at pixel $(h,w)$ and $t$ is the criticality threshold.

\subsection{Strategic Attack Generation for Robustness Testing}
The AEs integrate temporal class priority with spatial feature attribution for efficiency-optimized adversarial attack:
\begin{equation}
\label{attack}
\tilde{x}_i = x_i + \epsilon \cdot B_i \odot \text{sign}(\nabla_{x_i} J(x_i, y_{true}))
\end{equation}
where $\epsilon$ controls perturbation magnitude, $B$ restricts modifications to XAI-identified critical regions, $J(x_i, y_{true})$ represents the classification loss for ground truth $y_{true}$, and $\nabla_{x_i}$ denotes the gradient for input $x_i$. This dual-domain optimization achieves computational efficiency by attacking only priority classes (temporal efficiency) and only influential pixels (spatial efficiency), significantly reducing resource overhead while maintaining adversarial attack effectiveness.
% \vspace{-0.25cm}
\begin{algorithm}[htbp]
\caption{Dual-Domain Strategic Attack (DDSA) for Image Classification Processing}
\label{alg:ddsa}
\begin{algorithmic}[1]
\Require{Dataset $X$, attack threshold $\tau$, saliency threshold $t$, perturbation $\epsilon$}
\Ensure{Adversarially perturbed images $\{\tilde{x}_i\}$}
\For{each image $x_i \in X$}
    \State Predict class $\hat{y}_i$ and probability vector $\mathbf{p}_i$
    \State Compute trigger score $Q_i$ based on Eq. \eqref{q_function}
    
    \If{$Q_i > \tau$}
        \State Generate saliency map $L_i$ using IG (Eq. \eqref{s2})
        \State Create binary mask $B_i$ based on Eq. \eqref{s3}
        \State Generate adversarial example $\tilde{x}_i$ based on Eq. \eqref{attack}
    \Else
        \State $\tilde{x}_i = x_i$ \Comment{Skip attack, keep original image}
    \EndIf
\EndFor
\State \Return $\{\tilde{x}_i\}$
\end{algorithmic}
\end{algorithm}
% \vspace{-0.55cm}

\section{Experiments}
We conduct comprehensive simulation studies using diverse datasets to evaluate DDSA's dual-domain effectiveness. To validate our framework's performance, we design three complementary experiments: (1) spatial domain validation examining XAI-guided pixel targeting effectiveness and spatial coverage efficiency, (2) temporal domain assessment focusing on testing time savings through selective processing, and (3) Overall performance evaluation to assess resource allocation efficiency across priority classes. 

% \vspace{-0.35cm}
\subsection{Experimental Setup}
The experimental framework simulates real-world image transmission scenarios where computational resources are constrained and class priorities differ substantially. The simulation employs CIFAR \cite{krizhevsky2009learning}, Fashion-MNIST \cite{xiao2017fashion}, and VOC2012 \cite{everingham2010pascal} datasets with well-trained ResNet-18 classifiers \cite{he2016deep}, chosen for their varied class complexity enabling thorough evaluation across different classification challenges. The class diversity, ranging from 10-class scenarios to complex 100-class environments, allows us to simulate realistic deployment scenarios where content criticality varies significantly. For spatial domain evaluation, we employ FGSM and PGD variants (10/20/40/60 iterations) as baseline attack methods, comparing full-image attacks with spatially-targeted approaches using Integrated Gradients for pixel importance assessment. Spatial coverage is systematically varied from 100\% to 40\% to quantify efficiency gains. For temporal domain assessment, we measure execution times across different attack algorithms on large-scale dataset to evaluate computational savings. The overall performance implements class-priority scenarios reflecting practical applications: high-priority classes (humans, vehicles) receive maximum attention weights while routine objects (vegetation, background elements) are assigned lower priorities. The reward system assigns +1 for successful attacks on critical classes, -0.5 for failed attempts, and -0.3 for attacks on non-critical classes, enabling quantitative assessment of resource allocation efficiency. All experiments use ResNet-18 architectures trained to convergence on respective datasets, with consistent hyperparameters across comparative evaluations to ensure fair assessment of DDSA's overall effectiveness and efficiency.

% \vspace{-0.35cm}
\subsection{Spatial Domain Validation}
\begin{figure}[htbp]
\centering
\includegraphics[width=0.9\columnwidth]{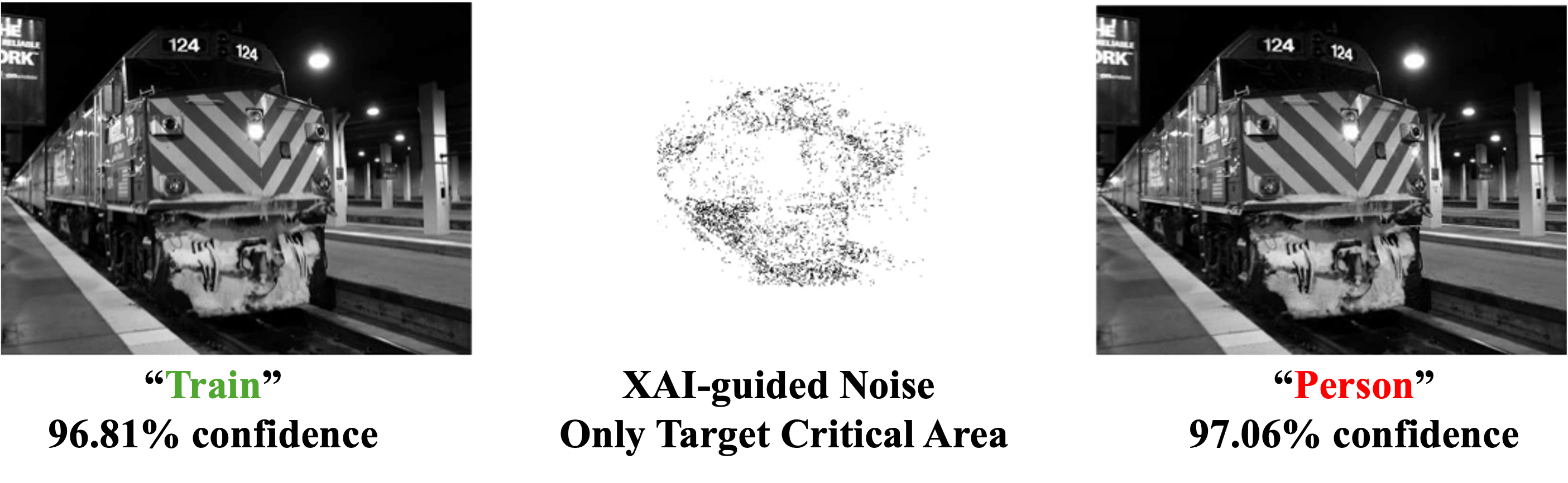}
% \vspace{-0.65cm}
\caption{XAI-guided spatial targeting demonstration.}
\label{fig:XAI_attack}
\end{figure}

\textbf{XAI enables precise spatial targeting.} To demonstrate the effectiveness of Integrated Gradients in identifying critical regions, we conducted targeted pixel perturbation experiments on high-quality images of VOC dataset. Fig.~\ref{fig:XAI_attack} shows a representative case where a train image originally classified with 96.81\% confidence is successfully misclassified as a person with 97.06\% confidence through targeting perturbation, validating that XAI can precisely locate critical pixels with maximum influence on classification decisions.

% \vspace{-0.4cm}
\begin{table}[htbp]
\centering
\small
\begin{tabular}{lccccc}
\toprule
\textbf{Method} & \textbf{100\%} & \textbf{90\%} & \textbf{80\%} & \textbf{60\%} & \textbf{40\%}\\
\midrule
\multicolumn{6}{c}{\textit{CIFAR-10}} \\
\midrule
FGSM   & 84.36\% & 84.01\%& 83.21\% & 80.88\% & 76.06\%\\
PGD-10 & 98.77\% & 98.63\%& 98.29\% & 96.82\% & 93.96\%\\
PGD-20 & 99.80\% & 99.73\% & 99.65\% & 99.08\% & 97.58\% \\
PGD-40 & 99.84\% & 99.81\% & 99.76\%& 99.47\% & 98.38\%\\
PGD-60 & 99.87\% & 99.86\% & 99.77\%& 99.51\% & 98.64\%\\
\midrule
\multicolumn{6}{c}{\textit{Fashion-MNIST}} \\
\midrule
FGSM   & 78.52\% & 78.42\% & 78.36\% & 78.14\% & 76.25\%\\
PGD-10 & 86.93\% & 86.11\% & 85.42\% & 83.53\% & 81.36\%\\
PGD-20 & 88.92\% & 88.33\% & 87.60\% & 85.76\% & 82.89\% \\
PGD-40 & 89.07\% & 88.49\% & 87.76\%& 85.88\% & 83.01\%\\
PGD-60 & 89.11\% & 88.5\% & 87.88\%& 85.89\% & 83.11\%\\
\bottomrule
\end{tabular}
\caption{Attack Success Rate vs. XAI-Guided Spatial Coverage}
\label{tab:ASR_combined}
\end{table}
% \vspace{-0.25cm}
% \textbf{Spatial reduction preserves effectiveness.} To quantify the relationship between the spatial coverage of XAI-identified critical regions and attack success rates, we simulated image processing systems across multiple datasets and systematically varied pixel coverage from 100\% down to 40\%. The results in Table \ref{tab:ASR_combined} demonstrate the effectiveness of critical pixel targeting, with attacks maintaining exceptional performance despite dramatic spatial reductions. PGD-based methods prove particularly robust, with PGD-60 preserving over 98\% success rates even when targeting only 40\% of image pixels, validating that XAI-guided identification of critical regions enables substantial spatial efficiency gains without compromising attack effectiveness. The consistency of these findings across both CIFAR-10 and Fashion-MNIST datasets confirms the generalizability of our approach across diverse classification scenarios.

\begin{figure}[htbp]
\centering
\includegraphics[width=0.95\columnwidth]{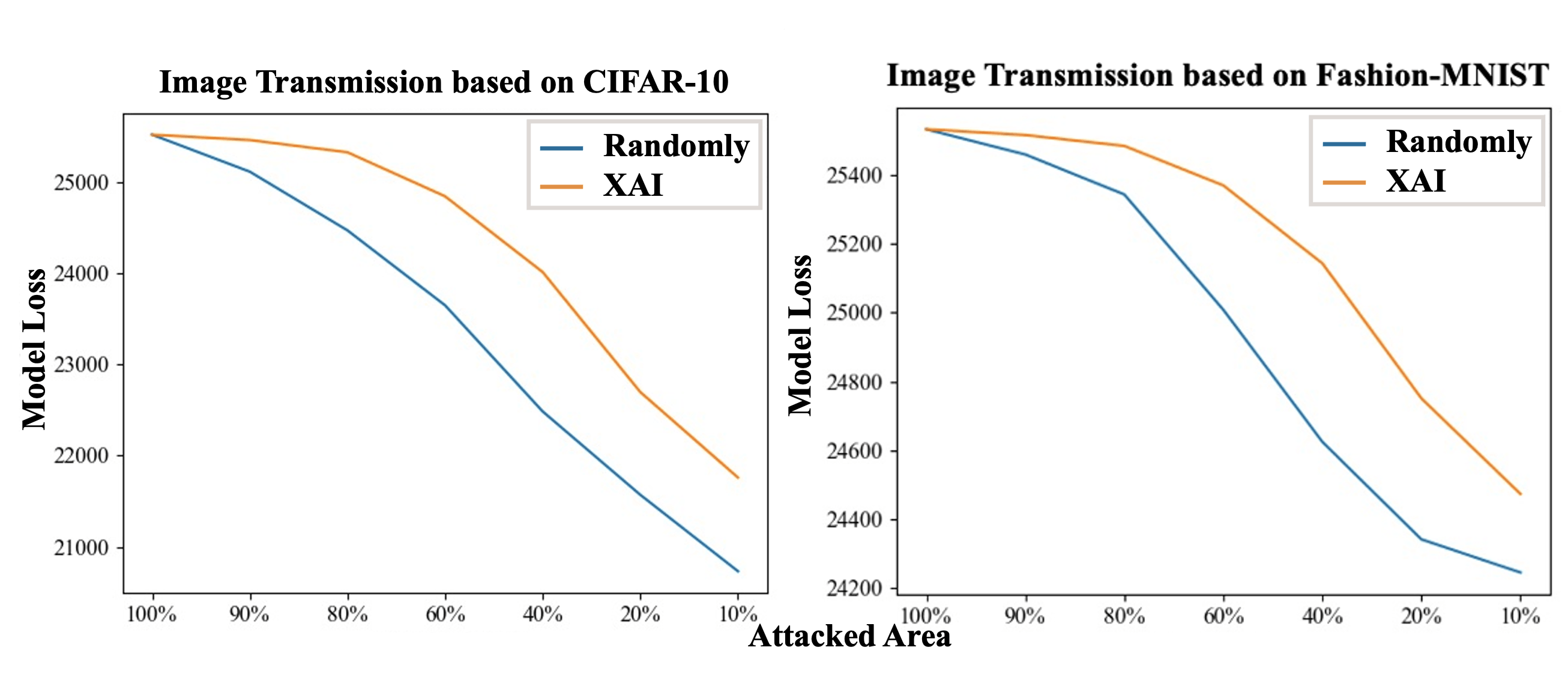}
% \vspace{-0.7cm}
\caption{Classifier loss under XAI-guided vs. random selection.}
\label{fig:loss_chart}
\end{figure}

% \textbf{XAI targeting outperforms random selection.} To validate the superiority of XAI-guided targeting over arbitrary pixel selection, we further conducted comparative analysis between XAI targeted pixel and random pixel across varying attack areas. Fig.~\ref{fig:loss_chart} presents empirical evidence that XAI pixel selection consistently achieves higher classifier loss compared to random perturbations, confirming that spatial effectiveness stems from principled feature attribution rather than arbitrary pixel reduction.
\textbf{Spatially efficient XAI-guided targeting preserves attack effectiveness.}
To quantify the relationship between spatial coverage of XAI-identified critical regions and attack success, we evaluate image classification systems across multiple datasets while systematically reducing the perturbed area from 100\% to 40\%. As shown in Table~\ref{tab:ASR_combined}, attacks maintain exceptionally high success rates despite dramatic spatial reduction, demonstrating the effectiveness of targeting XAI-identified pixels. PGD-based methods are particularly robust, with PGD-60 preserving over 98\% success even when perturbing only 40\% of image pixels. Moreover, comparative analysis against random pixel selection (Fig.~\ref{fig:loss_chart}) shows that XAI-guided perturbations consistently induce higher classifier loss across all attack areas, confirming that the observed spatial efficiency arises from principled feature attribution rather than arbitrary pixel reduction. The consistency of these results across both CIFAR-10 and Fashion-MNIST further validates the generalizability of our approach across diverse classification scenarios.

% \vspace{-0.65cm}

\begin{figure}[htbp]
\centering
\includegraphics[width=0.9\columnwidth]{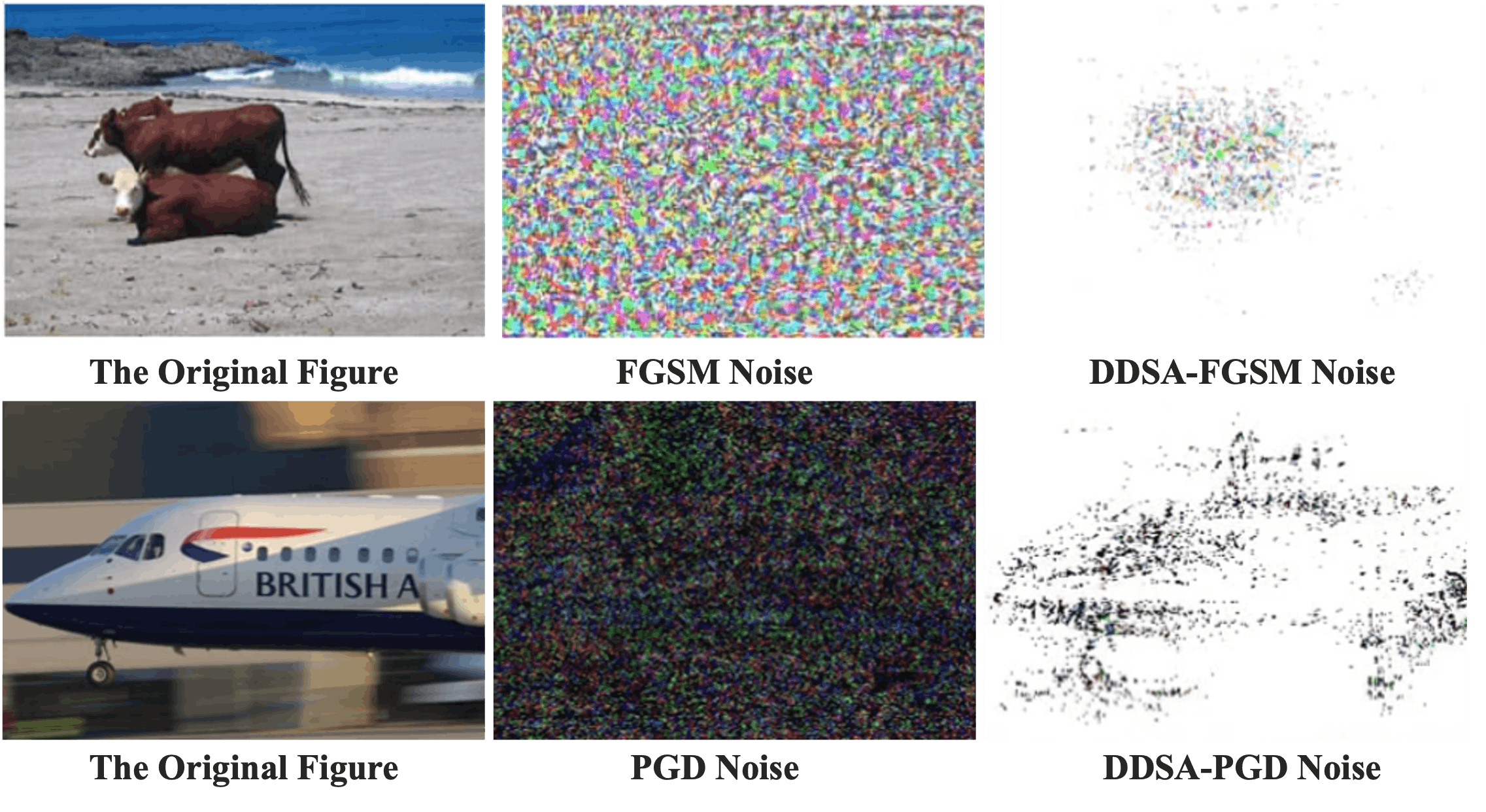}
% \vspace{-0.6cm}
\caption{Testing perturbation comparison: full-image vs. DDSA.}
\label{fig:noise_comparison}
\end{figure}

% \vspace{-0.15cm}
\textbf{Minimal modifications achieve maximum efficiency.} To assess the overall spatial advantage of our approach, we analyzed pixel modification requirements across entire simulation. Comprehensive analysis reveals strategic attacks require targeting merely \textbf{\textit{30\% of pixel area to achieve 95\% success rates, affecting only 0.0123\% of total pixels simulation-wide}}. To visualize this spatial efficiency advantage, we generated comparative noise patterns shown in Fig.~\ref{fig:noise_comparison}, demonstrating substantially reduced visual noise compared to traditional full-image robustness testing approaches.

% \vspace{-0.4cm}
\subsection{Temporal Domain Assessment}
\begin{figure}[htbp]
\centering
\includegraphics[width=0.9\columnwidth]{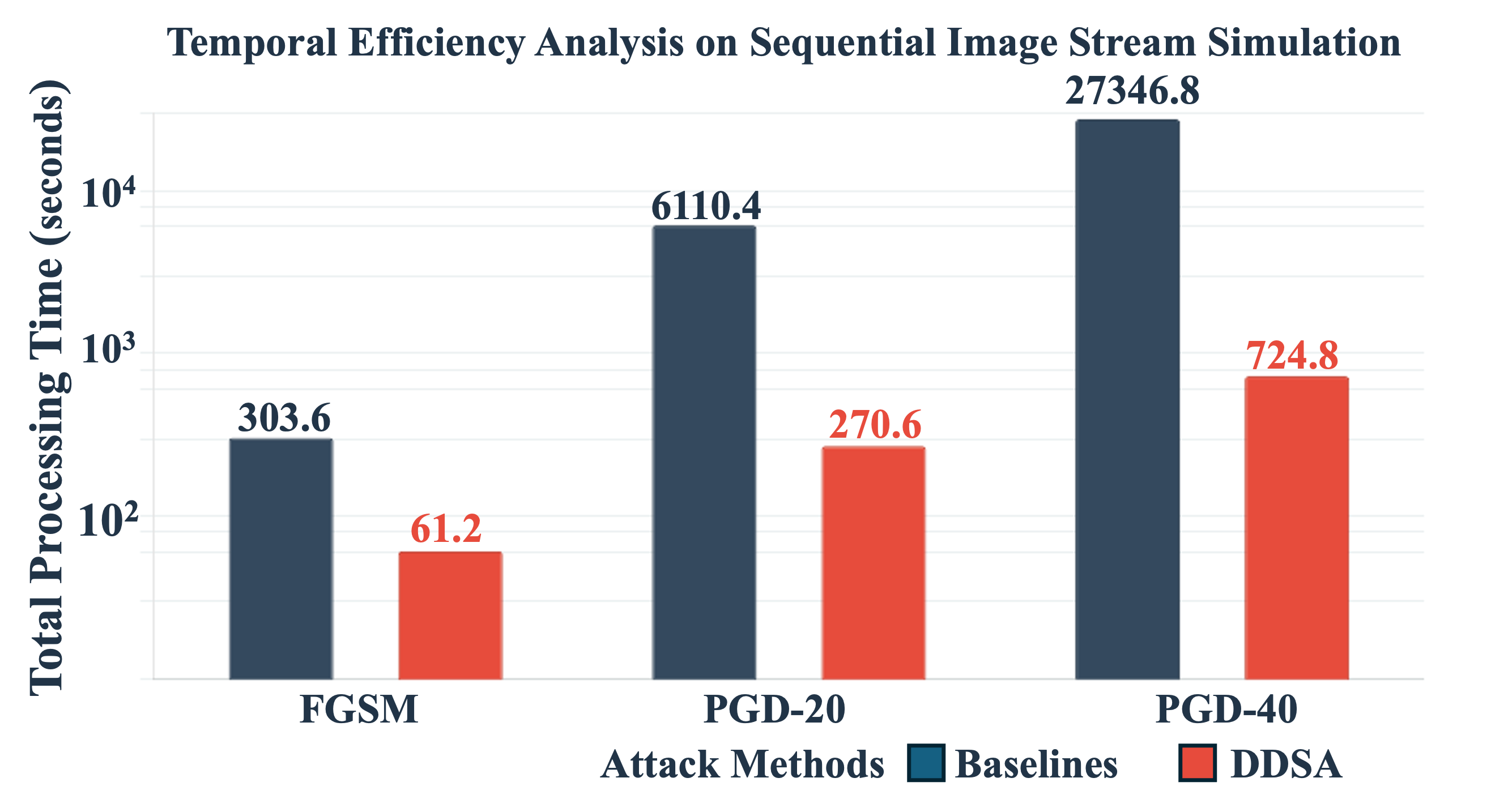}
% \vspace{-0.6cm}
\caption{DDSA achieves 80-97\% computational time reduction.}
\label{time_analysis}
\end{figure}

\textbf{Temporal selectivity achieves significant speedups.}
To quantify the efficiency gains from temporal selectivity, we measure testing times across multiple attack configurations, explicitly accounting for the overhead of saliency map generation. As shown in Fig.~\ref{time_analysis}, DDSA substantially reduces evaluation time by mitigating a key inefficiency in conventional adversarial testing, which indiscriminately processes entire image sequences regardless of their semantic relevance. In realistic continuous monitoring scenarios, most frames contain non-critical content that poses minimal risk if misclassified, yet are redundantly subjected to exhaustive testing. By leveraging a trigger function to identify and prioritize mission-critical frames, DDSA enables targeted robustness evaluation that focuses on consequential content while avoiding unnecessary computation. This temporal filtering strategy makes comprehensive adversarial testing feasible for large-scale, continuous image streams, directly addressing the scalability challenge of robustness evaluation in real-time deployments.

% \textbf{Temporal selectivity achieves significant speedups.} To evaluate the computational benefits of temporal aspect, we measured testing times across multiple attack algorithms and datasets, including the overhead introduced by saliency map generation. Our approach mitigates a key inefficiency in general adversarial attacking, which indiscriminately processes entire image sequences regardless of their critical content. In real-world deployments such as continuous image monitoring, most frames contain non-critical objects that pose minimal risks if misclassified, yet traditional methods redundantly subject every frame to exhaustive testing.

% The observed efficiency in Fig. \ref{time_analysis} gains highlight the practical value of intelligent temporal selectivity in large-scale deployments where time and computational resources are critical. By employing our trigger function to identify and prioritize only mission-critical frames, DDSA enables targeted robustness evaluation that focuses on genuinely consequential content rather than routine imagery. This strategy makes comprehensive adversarial testing computationally feasible for continuous large-scaled image streams, addressing the issus of scaling robustness evaluation to real-time applications.

% \vspace{-0.4cm}
\subsection{Large-scale Image Stream Processing Simulation}
\textbf{DDSA enables massive image streams processing.} To evaluate DDSA's overall effectiveness in high-throughput scenarios, we simulated large-scale image stream processing systems, processing continuous streams of 30,000 images with distributions randomly sampled from the aforementioned datasets, where critical classes (4 predefind classes per dataset to emulate mission-critical objects such as vehicles and humans) require immediate detection while routine imagery is deemed non-essential for mission objectives.

Table~\ref{tab:realtime_simulation} demonstrates DDSA's superior performance in high-throughput scenarios, with all DDSA variants achieving positive operational scores across different complexity levels. Conventional approaches behave inefficient in complex scenarios, yielding severe negative scores due to indiscriminate processing of non-critical content. This validates that temporal and spatial selectivity are essential for deploying adversarial robustness testing in real-time applications where computational efficiency directly impacts mission success.

\begin{table}[htbp]
\centering
\small
\begin{tabular}{lccc}
\toprule
Method & CIFAR-100 & Fashion-MNIST & CIFAR-10\\
\midrule
FGSM & -2535.5 & 1054 & 1492\\
DDSA-FGSM & \textbf{353} & \textbf{3556} & \textbf{3462}\\
PGD-20 & -2481.5 & 1562.5 & 2186.5\\
DDSA-PGD-20 & \textbf{385} & \textbf{3331} & \textbf{3986}\\
PGD-60 & -2475.5 & 1574.5 & 2197.5\\
DDSA-PGD-60 & \textbf{388} & \textbf{3832} & \textbf{3993}\\
\bottomrule
\end{tabular}
% \vspace{-0.2cm}
\caption{Simulation results across different datasets.}
\label{tab:realtime_simulation}
\end{table}

% \vspace{-0.45cm}
\section{Conclusion}
This paper presents DDSA, a dual-domain framework that achieves superior computational time reduction and targeted processing of only critical pixels while maintaining high attack success rates in adversarial robustness testing through temporal selectivity and spatial precision. Our approach enables practical deployment of comprehensive robustness testing in resource-constrained real-time applications, addressing the critical scalability gap between laboratory effectiveness and real-world deployment.

\clearpage
\section{Acknowledgments}
This work is partially funded by the European Union (Grant Agreement No. 101212818). The views and opinions expressed are those of the authors and do not necessarily reflect those of the European Union or the European Health and Digital Executive Agency (HADEA). This work is also partially supported by Innovate UK through the AI-PASSPORT programme (Grant No. 10126404).
% Yi Dong’s contribution is partially supported by the Royal Society International Exchanges Programme and the Engineering and Physical Sciences Research Council (EPSRC) through RAi UK (Grant No. EP/Y009800/1).
% This work is partially funded by the European Union (under grant agreement ID 101212818). Views and opinions expressed are however those of the author(s) only and do not necessarily reflect those of the European Union or European Health and Digital Executive Agency (HADEA). Neither the European Union nor the granting authority can be held responsible for them. 
% This work is partially supported by Innovate UK through AI-PASSPORT under Grant 10126404. 
% Yi's contribution is partially supported through the Royal Society international exchanges programme and in part by the Engineering and Physical Sciences Research Council, through funding from RAi UK [EP/Y009800/1].

% \bibliographystyle{IEEEbib}
\bibliographystyle{ieeetr}
% {\footnotesize
\bibliography{strings,refs}
% }
\end{document}